# Risk-Aware Fine-Grained Access Control in Cyber-Physical Contexts

JINXIN LIU, University of Ottawa, Canada
MURAT SIMSEK, University of Ottawa, Canada
BURAK KANTARCI, University of Ottawa, Canada
MELIKE EROL-KANTARCI, University of Ottawa, Canada
ANDREW MALTON, BlackBerry Ltd., Canada
ANDREW WALENSTEIN, BlackBerry Ltd., Canada

Access to resources by users may need to be granted only upon certain conditions and contexts, perhaps particularly in cyber-physical settings. Unfortunately, creating and modifying context-sensitive access control solutions in dynamic environments creates ongoing challenges to manage the authorization contexts. This paper proposes RASA, a context-sensitive access authorization approach and mechanism leveraging unsupervised machine learning to automatically infer risk-based authorization decision boundaries. We explore RASA in a healthcare usage environment, wherein cyber and physical conditions create context-specific risks for protecting private health information. The risk levels are associated with access control decisions recommended by a security policy. A coupling method is introduced to track coexistence of the objects within context using frequency and duration of coexistence, and these are clustered to reveal sets of actions with common risk levels; these are used to create authorization decision boundaries. In addition, we propose a method for assessing the risk level and labelling the clusters with respect to their corresponding risk levels. We evaluate the promise of RASA-generated policies against a heuristic rule-based policy. By employing three different coupling features (frequency-based, duration-based, and combined features), the decisions of the unsupervised method and that of the policy are more than 99% consistent.

CCS Concepts: • **Security and privacy** → **Systems security**; **Software and application security**; **Access control**; **Authorization**; Authentication; • **Computing methodologies** → *Machine learning*.

Additional Key Words and Phrases: Cyber-security, IoT, Access Control, Authentication, Machine learning

## 1 INTRODUCTION

Whether access should be granted to a resource may depend upon the context in which the access is attempted, including the cyberphysical context. For example, a physician may have a legitimate need to access a patient's private health information (PHI) on a mobile display screen while in a private consultation room, but it might be important to lock out the access if the physician wheels the display into the hallway, or if a different patient is moved into the shared room. Failure to lock out could lead to leaking PHI to other people who the patient has not consented to allow access to. In such cases the identity of the accessing physician has not changed, nor

Authors' addresses: Jinxin Liu, jliu@uottawa.ca, University of Ottawa, Ottawa, ON, K1N 6N5, Canada; Murat Simsek, murat.simsek@uottawa.ca, University of Ottawa, Ottawa, ON, K1N 6N5, Canada; Burak Kantarci, burak.kantarci@uottawa.ca, University of Ottawa, Ottawa, ON, K1N 6N5, Canada; Melike Erol-Kantarci, melike.erolkantarci@uOttawa.ca, University of Ottawa, Ottawa, ON, K1N 6N5, Canada; Andrew Malton, amalton@blackberry.com, BlackBerry Ltd., Waterloo, ON, K1N 6N5, Canada; Andrew Walenstein, awalenstein@blackberry.com, BlackBerry Ltd., Waterloo, ON, K1N 6N5, Canada.

has the physician's *role* changed with respect to the patient. The concern is thus unrelated to *authentication* and static user *roles*. Instead, it is an issue of context-dependent or context-sensitive *authorization*.

Recently, context-aware security approaches have gained significant attention, and there have been several studies around inferring and understanding contexts. But even state-of-the-art access control mechanisms for cyber-physical systems may lack the generality or flexibility desired to confine access authorization to dynamic cyberphysical context, or they otherwise burden the access policy administrators with having to tediously and continuously define every context, set the appropriate access restrictions, and understand the resulting policy well enough to assess the residual risks left by the policy. For example, methods for defining *geofences* may be used to define physical contexts for defining context-dependent access policy, but many geofencing access policies utilize geofences only for authorization decisions, and physical location is one contextual factor affecting authorization risks. For example, a phone may lock when moved out of its "home" geofence, but unlocking restores all prior access, it is only used for context-dependent authentication.

A general framework and approach is required for effective context-dependent authorization policy management. Core concerns in any such context-dependent authorization regime is defining the contexts in which authorization decisions differ, identifying and quantifying risks for those contexts, and then using the derived risk model to create appropriate access control decisions. Given that contexts may change over time, an additional burden is tracking these changing contexts over time. A problem that is not yet well addressed are effective methods for automatically helping define and maintain context-dependent authorization policies, i.e., inferring the contexts, risk models, and access decision boundaries.

This paper introduces a framework named Risk-Aware Smart Access (RASA) to learn authentication policies from examples of routine access. RASA begins with the observation that *if*, in normal practice, responsible actors access information they have ordinary access to only in those contexts in which the need access, and that they take care to guard against known risks. In such cases, even if an existing access control policy is not context-sensitive, responsible actors effectively fill in the missing contextual access conditions through their choices and actions, and these can be used to infer a baseline context-dependent access control policy. In a sense, they can program the context-dependent access conditions by demonstration. RASA uses coupling of access actions to identify patterns of access, clusters the actions, associates these coupling-defined features to risk scores, and then defines access boundaries based simply on these risk scores. To the best of our knowledge, no prior work exists for similar automated policy inference based on mapping inferred contexts onto risk levels.

The paper evaluates the general promise of the approach in a simulated in-person healthcare environment. This use context is cyberphysical, and one where ensuring strong access boundaries–even for fully authenticated users–are critical for protecting sensitive PHI. Nearly 90% of participants, in a health care survey reported by the Ponemon Institute, have been affected by data leakage in the past three years [22]. Furthermore, inappropriate use of devices that handle sensitive document by internal employees has led to the breach of 4.5M patient records in 2015 [14]. Intelligent approaches to such security may be increasingly important [30], and not all risks are due to remote attackers–shoulder surfing in a shared patient room, for example. The context-dependent appropriateness of PHI access serves as a suitable test of cyberphysical context dependencies.

The paper is organized as follows. In Section 2, the related work and motivation are presented. In Section 3, we present our proposed framework in detail. Section 4 presents the numerical results of three different coupling mechanisms and clustering algorithms along with comprehensive discussions. Finally Section 5 concludes the paper and provides future directions.

## 2 BACKGROUND, RELATED WORK, AND MOTIVATION

Some traditional authentication schemes, such as passwords or pincodes, can be inconvenient for some users. They also have well-known limitations to their effective security; for example, passwords set by mobile users can be determined by simple guessing [1], inferred from smudges left on users' screens [5], or stolen by malware [6]. These concerns have lead to plentiful prior work in cyberphysical context-dependencies for access control that has applied context dependency to multi-factor authentication, location-aware access control, and implicit or continuous authentication.

Multi-factor authentication establishes likely identity of users based on at least two pieces of evidences of identity, making identity assessment dependent upon multiple possibly-contextual factors. Commonly, the factors relate to a users' unique knowledge (such as a private password), and physical possessions (something that is only in possession of the user such as a SecureID token). Common also is adding physical location to the authentication through definitions of geofences, such as a defined "home" location. Ramatsakane et al. [33] utilized location for improving accuracy of identity determination for authentication purposes.

So-called "implicit" authentication extends multi-factor one-time authentication could be more usable and secure [12, 24]. State of the art in context-aware access control calls for a system for continuous authentication that factors contextual risk into the continuous authentication in addition to other risks, such as the risk that the user is not who they appear to be, or that the resource being accessed is more sensitive than another [31]. Thanks to the rapid advancement of the Internet of Things (IoT), sensors can be utilized for easy-to-use additional access safeguards, such as through contextual, behavioral, and biometric signatures [32]. In particular, Internet of Things (IoT) devices and networks may be effective in recognizing certain context, as in the case of using sensors, communication and data (i.e. storage and analytics) as keys to IoT-based smart systems [16]. As Habibzadeh et al. [17] note, security is increasingly critical in smart access systems. Significant work on behavioural authentication has been published recently using a growing list of features collected through sensors including, but not limited to: smartphone touchscreens, wearables, and keystrokes [9, 38, 39].

Ashibani et al. [3] propose a framework which aims to reduce user intervention in access control by identifying the behavioral patterns about user-device interaction, such as service request and login duration. Rule-based behavior analysis aims for resiliency against password misuse, brute force attacks and unauthorized modification. Rauen et al. [35] propose a primary-backup fashion to manage access control on smart mobile devices by using several machine learning algorithms that run on gestural information and spatiotemporal knowledge extracted from sessions on various applications. Thus, once the primary authentication module fails to authenticate a user (or fails to recognize a behavioral pattern), the backup module is activated. In case both primary and backup modules fails to authenticate a user implicitly, multi-factor authentication is triggered. Wu et.al [42] apply Support Vector Machines (SVMs) to classify genuine users as well as adversaries from keystroke signals. Lima et al. [28] introduced an architecture for behavioural data acquisition, and presented a behavior model with the building blocks of events, context, action and behavior. The architecture utilizes a belief analyzer and a recommendation filter to discover anomalous behaviour, and uncover hidden behavioural patterns. In addition to these, a probability analyzer is proposed to serve as a classifier that outputs three categories: normal, suspicious and abnormal. Hayashi et al. [18] proposed a framework named "Context-Aware Scalable Authentication" (CASA) to make a trade-off between security and usability. In addition to multiple factors that contribute to a context, the authors report that location is the most influential factor on the context. Lee [27], a new continuous authentication system called "SmarterYou" is introduced to integrate the data from smart phone with wearable sensors. The proposed system builds on data analytics on these features in both time and frequency domains so to ensure fine-grained authentication. Flexibility is another important aspect to address in smart access control research as it can potentially compromise security. To address this challenge, Hulsebosch et al. [21] provide a secure and flexible access control scheme which builds on context sensitiveness and historical data analysis in the authorization of anonymous users.

While techniques to recognize behavioral patterns have improved, behavioral authentication still cannot guarantee the complete elimination of false positives [2]. In addition, even if false positives can be solved for authentication, implicit authentication alone cannot solve context-dependent risks that may result in undesired access.

Table 1. Access Control Schemes Comparison

| Access control scheme | Benefits | Limitation |
|---|---|---|
| MAC | High security level; Fine-grained; Attackers cannot share or inherit access to other files | Difficult to maintain and configure; Difficult to scale to a large number of files and users; Users have to request permissions for new files |
| DAC | Easy to maintain and configure; User friendly; Low implementation cost; Flexible | Less secure than MAC; Lack of centralized access management |
| RBAC | Rules are transparent to users; Ease of use for the administrator; Flexible; Secure but no need to configure access for each user | Complex and laborious configuration of roles; Difficult to extend permission for individual users; Complexity is determined by the number of roles; Discards the principle of least privilege |
| ABAC | Dynamically updates access permissions; Less effort after configuration; More control variables than RBAC | Low interpretability for the user; Scalability issues |
| Proposed Scheme (RASA) | No configuration; Infer document ownerships automatically; Dynamically adjust access permission according to user's environment; Identify anomalous actions automatically | Sensitive to surroundings. Not suitable for small number of users since it is difficult to infer the coupling distributions |

Access control models such as Mandatory Access Control (MAC), Discretionary Access Control (DAC), Role-Based Access Control (RBAC), and Attribute-Based Access Control (ABAC) have been investigated by the researchers in this domain [11]. MAC offers operating system-constrained access control to manage process/thread operating files, network ports, memories, and devices. MAC is widely deployed on Linux, Windows, and databases. Rossi *et al.* [37] propose a framework that allows developers to establish fine-grained ad-hoc MAC strategies for applications, protecting the system from misbehavior (caused by bugs or compromised by attackers) of root privileged apps. Even though MAC provides high-level protection, it requires users to request permission for every resource. When compared to MAC, DAC is more flexible by allowing users to grant access to other users. Khan *et al.* [25] apply DAC combined with RBAC to access patient files dynamically. RBAC, unlike MAC and DAC, does not assign access to specific users but to pre-defined roles. By deploying RBAC, IT administration can operate more efficiently since RBAC can add or switch roles to cope with personnel changes. Cruz *et al.* [8] propose RBAC-SC, which utilizes RBAC and smart contracts to implement trans-organizational roles. Unlike static access control strategies, ABAC dynamically determines permissions according to a set of attributes such as user, resource, object, environment attributes. ABAC can achieve various levels of access control according to different requirements and environments in spite of its overhead. Ding *et al.* [10] propose combining ABAC and blockchain to cope with the massive connectivity of IoT so to avoid heavy role-engineering for various devices. As listed in Table 1, compared to the conventional access control schemes, this paper focuses on implicit access control, which can identify the user-object association by mining historical data, and recognize risky actions according to the current context.

Besides the studies that aim to bridge access control with implicit authentication, identifying and assessing the risks of certain contexts is of paramount importance as well. With this in mind, the authors in [19] define a risk assessment scheme that associates the risk to the location. Cha *et al.* [7] introduce a data-driven risk assessment scheme that uses Data Flow Diagrams (DFDs) to model the flows of individual data and to recognize the components employed to process, store, and transmit data. Organizations can identify the potential/implicit incidences according to the associated components. Compared with asset-oriented and process-oriented approaches, the proposed approach can enhance the risk assessment accuracy by avoiding underestimating or neglecting the risks to sensitive individual data. In [23], Khambhammettu *et al.* propose a framework containing four threat assessment methods for subject-object accesses. Via assigning different weights to the sensitivity score of data/objects and the trustworthiness score of subjects, the proposed framework is flexible enough to satisfy the various preference of organizations. Wang *et al.* [40] present a new access control model that quantifies the risk of privacy violation in a statistical approach and further detects the physicians/users who over-access or misuse patients' private data. Atlam *et al.* [4] propose a risk estimation model that combines a new fuzzy logic algorithm and a set of rule-based policies to perform access control in IoT systems. To generate accurate and realistic risk values for each access request, the proposed fuzzy logic system, which involves twenty experts' effort, converts experts' qualitative expression into numeric values and offers a dynamic and context-aware access control by leveraging the contextual features such as resource sensitivity, action severity, and risk history.

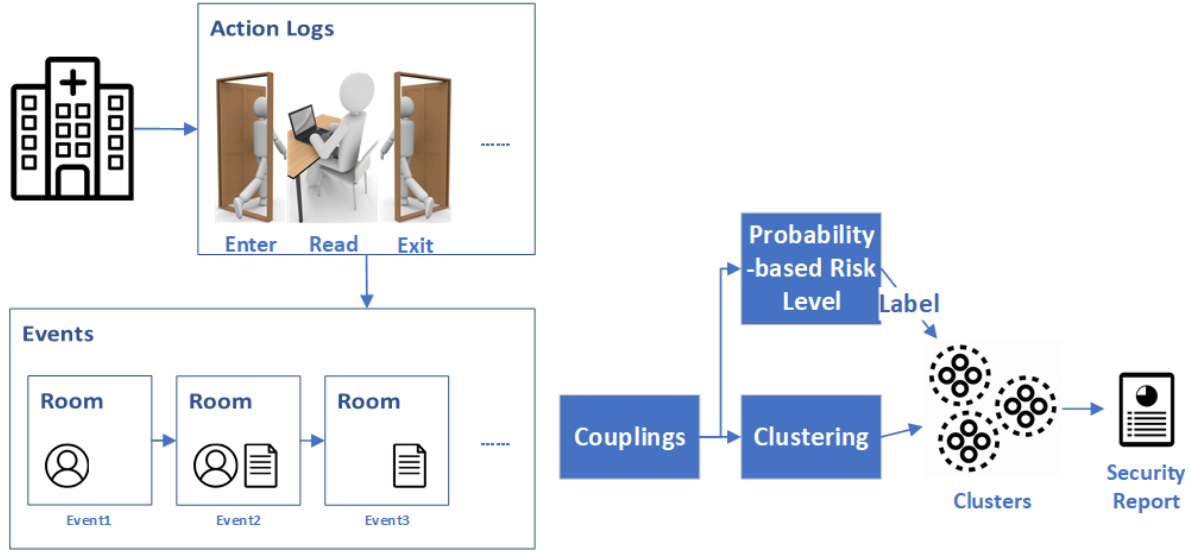

Fig. 1. Overview of System Architecture

The literature review indicates a gap remains between continuous/implicit authentication/access control and risk-awareness, particularly in cyberphysical environments where actions do not only appear in cyberspace but also stem from physical behaviors, interactions and roles.

## 3 RASA: RISK-AWARE SMART ACCESS CONTROL

This section presents the system overview along with the cyber-physical action logs in a medical emergency room. We introduce the data pre-processing steps, feature extraction and risk inference methods in detail.

Table 2. Action Log Examples

| time | act | agent | device | document | location | monitor |
|---|---|---|---|---|---|---|
| 06-26 19:00 | enter | actor: 35...9d | | | loc: fe...fb | dev: bd...77 |
| 06-26 20:02 | exit | actor: 50...df | | | loc: fe...fb | dev: bd...77 |
| 06-26 19:31 | read | | dev: ae...2e | doc: 9a...de | | |

In our framework, actions are transformed into events which include contextual information that can be recognized through coupling mechanisms. Coupling defines interaction patterns between the objects, which can be recognized through clustering algorithms, that use couplings as the input features. Each coupling feature is associated to a risk level. Upon obtaining the clusters, an aggregate value of the associated risk levels of the features of all data points (i.e. actions) in a cluster is calculated as the risk value of that cluster, (i.e. label of cluster). Note that couplings can be formulated based on time, frequency or a combination of time and frequency.

The bold letters in the notation represent collections/sets of factors ($\boldsymbol{Ppl}$, $\boldsymbol{Dev}$, $\boldsymbol{Doc}$, and $\boldsymbol{Loc}$), and lower case letters stand for the elements of the specific collection, e.g. $\boldsymbol{A} = \{a_1, ..., a_i, ..\}$. $\boldsymbol{A}$ and $\boldsymbol{B}$ denote any pair of the four collections/sets. $C$ and $R$ indicate coupling and risk respectively and can be denoted in terms of Frequency ($Freq$) and/or Duration ($Dur$). $F_n^{A,B}$ denotes the $n^{th}$ coupling feature extracted (from the $n^{th}$ event ($E_n$)). More detailed explanations are listed in Table.3.

## 3.1 System Overview

The overview of the proposed RASA scheme is presented in Fig. 1. The use case in this study involves an emergency room of a healthcare organization where sensors record activities of people (i.e., patients and physicians) and devices, such as entering or exiting some locations and reading documents/records (as examples in Table 2). In order to protect patient privacy and physical security, protection systems should raise alarms or introduce access control levels against hazardous actions, such as a read access to a document by an unfamiliar patient or entering into unfamiliar rooms. The proposed scheme aims to quantify the risk of given actions and find potential threats without prior knowledge (e.g., document ownership, patient wards). Nevertheless, if only the action were considered, not only the context information would be overlooked but similar actions would be repeated, which might have led to many false alarms. Therefore, in the proposed framework, the first step is to reproduce the events where actions are performed. Then, these events are used for calculating the couplings which are then clustered and at the same time assigned some risk levels. The risk levels of couplings determine the risk level of a cluster of couplings. Finally, the risk levels of clusters are verified against a pre-defined policy.

An argument for using couplings as features described is as follows. Since the risk level is not only determined by actions but also their context, the actions are initially transformed into events to obtain the surrounding information, e.g. location, and co-presence. The coupling concept (see Section 3.3 for more details) is introduced to infer the relationship of objects in action logs. Coupling values are extracted from events and normalized. These values are stored in coupling matrices to explore their distributions and are marked with high, medium, and low-risk levels according to their mean value and standard deviation. Coupling features are fed into clustering algorithms. In this paper, three different coupling features are used: frequency-based features, duration-based features, and a combination of them. A simple scenario that would trigger the RASA framework to compute coupling values is illustrated in Fig. 4. One physical action of entering a room and one cyber action of accessing a document on a device is given as an example.

After the computation of couplings, their risk levels, and forming clusters of couplings, the risk of each cluster is calculated based on the number of high, medium and low labels of couplings in that cluster. Then, the cluster is labeled according to cluster risk value. Furthermore, the risk of the entire action log can be determined by the same calculation to provide the general risk information to the security experts to raise alarm.

Table 3. Notations

| Notation | Description |
| --- | --- |
| $C^{Freq}_{a_i,b_j}$ | Frequency-based Normalized Coupling Values |
| $C^{Dur}_{a_i,b_j}$ | Duration-based Normalized Coupling Values |
| $R^{Freq}_{a_i,b_j}$ | Risk of Frequency-based Normalized Coupling Values |
| $R^{Dur}_{a_i,b_j}$ | Risk of Duration-based Normalized Coupling Values |
| $Freq_{a_i,b_j}$ | Number of times $a_i$, and $b_j$ occur together |
| $Dur_{a_i,b_j}$ | Duration of the occurrence of $a_i$, and $b_j$ |
| $\delta_{a_i,b_j}(k)$ | Coupling distribution |
| $\boldsymbol{Ppl}$ | Collection of people |
| $\boldsymbol{Dev}$ | Collection of devices |
| $\boldsymbol{Doc}$ | Collection of documents |
| $\boldsymbol{Loc}$ | Collection of locations |
| $A, B$ | Any two of four collections, $\boldsymbol{Ppl}$, $\boldsymbol{Dev}$, $\boldsymbol{Doc}$, or $\boldsymbol{Loc}$ |
| $a_i, b_j$ | Elements from four collections |
| $E_k$ | The $k^{th}$ event which is defined as a collection of the elements from various set |
| $F^{A,B}_n$ | The coupling Feature |

In a privacy-sensitive environment such as a clinical setting, the inferred risk context can be utilized to define a rule-based access control policy managing access and defining risk-based system actions based on calculated risk. Given a risk level of a cluster of actions or events, a policy can be defined wherein access decisions are made per read attempt by calculating the overall risk within the current context. RASA estimates overall risk from the weighted sum of all per-feature risk values. Rather than assigning equal weights to the features, the policy author may adjust the weights of these coupling features based on direct inputs from subject matter experts. For instance, when patient documents are the assets considered, the couplings related to documents may be assigned greater weight factors compared to other couplings that do not consider documents. Furthermore, the policy may permit or deny read actions by tuning a threshold value for the risk factor. Although such policy definitions require a threshold to be tuned, the decisions are over clusters rather than individual actions, which is more scalable, stable, and explainable for policy analysts and users alike. In the conclusion part, we elaborate on AI outcomes versus policy decisions in order to analyze the efficiency of the AI results.

In the following sections, we explain the components of RASA in detail.

## 3.2 Action logs

In RASA's model, action logs consists on data of physical activity such as entering or exiting a room as well as cyber activity such as logging in a device or accessing a document. In a multi-user distributed system, the access control of documents is effective to all kinds of actions. Regardless of the devices used in a environment, wired or mobile, the access of a document can always be abstracted in association with user movements such as entering or exiting a room, device usage, and operating documents. Our simulated action logs are inspired by typical actors in clinical, in-person settings.

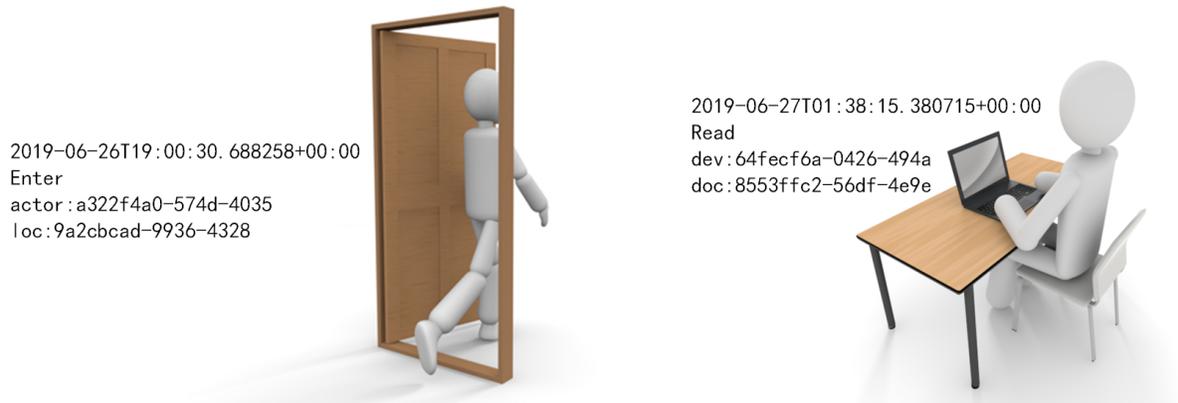

Fig. 2. Actions Examples

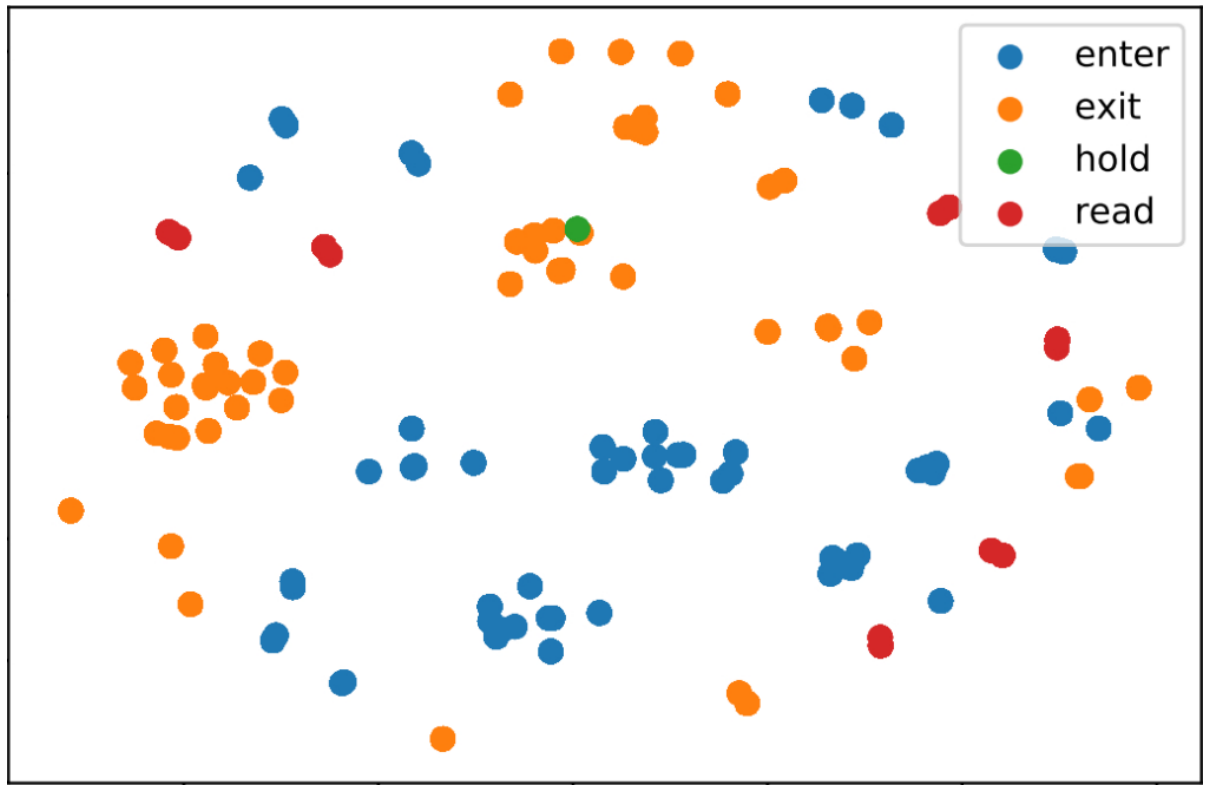

Fig. 3. Visualization Result of Raw Action log using t-SNE

An example of sequence of actions is given in Fig. 2. When users or devices enter or exit a specific location, the ID of the room and the ID of the user are logged with a timestamp. If a user attempts to access a document, the device first requests grant for access, and this attempt is also logged. The distribution of the simulated actions are visualized in Fig. 3 by using T-distributed Stochastic Neighbor Embedding (t-SNE).

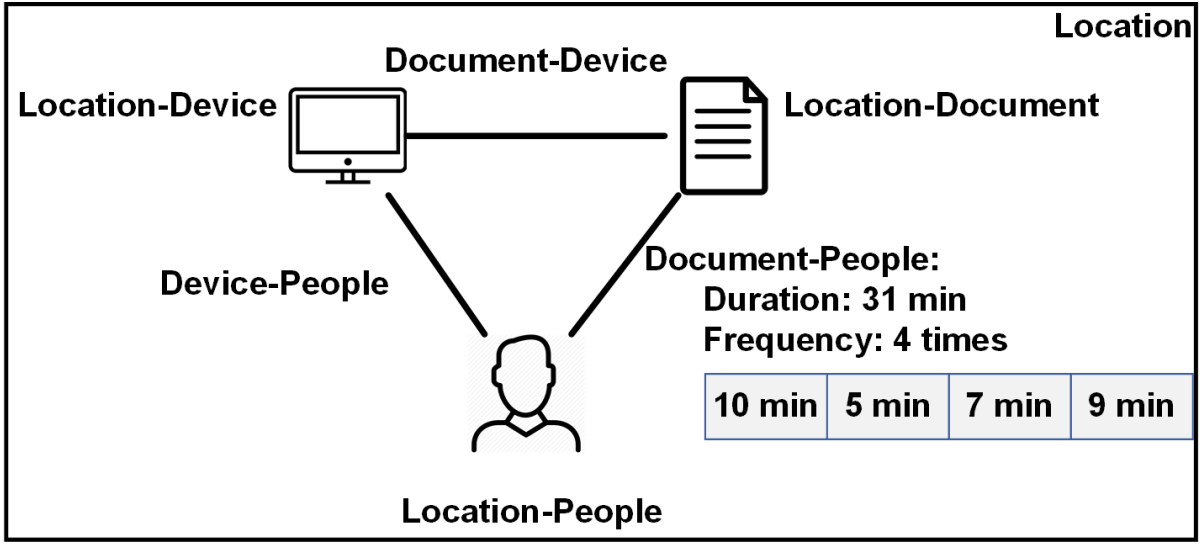

Fig. 4. A simple scenario to illustrate the coupling concept

As a widely used and effective statistical visualization technique, mapping high dimensional data onto 2D or 3D, t-SNE calculates the probability distributions of each pair of data samples in high and low dimensional spaces, assigning high probabilities to similar ones and minimizes two distributions Kullback–Leibler divergence. t-SNE can exhibit the data distribution lying in various manifolds and clusters, while the distance between two samples in low dimensional space do not represent the distance in high dimensional space but the probabilities [29]. Hence, the axes in the plot do not represent physical units. By visualizing the dataset with t-SNE, we aim to understand the dataset distribution, discover potential clusters, and analyze the complexity of the problems.

General characteristic of the actions can be formulated as moving these objects (i.e. people, documents and devices) across different locations over time. Thus, an event or a state machine is defined as a combination of time, location and a set of objects in this space.

### 3.3 Feature Extraction and Risk Inference

The coupling concept aims to reveal the interactions among different factors of action logs. As illustrated in Fig. 4, coupling can be defined between people and location, location and device as well as location and document. In the figure, the interaction between two factors: document and people is presented in terms of duration and frequency. Duration denotes how long two objects spend time with each other; frequency denotes how many times two elements encounter each other. Using the logs, RASA first constructs the frequency and duration matrices for each object pair (i.e. person-person, person-device, device-location, person-location). The coupling values in the matrices are normalized to enable comparability between different coupling values.

The coupling factors are denoted as: $Ppl = \{p_1, p_2, p_3, ..., p_n\}$, $Doc = \{doc_1, doc_2, doc_3, ..., doc_n\}$, $Dev = \{dev_1, dev_2, dev_3, ..., dev_n\}$, $Loc = \{loc_1, loc_2, loc_3, ..., loc_N\}$. The computation of coupling values is given in Eq.1 and Eq.2 for frequency and duration, respectively.

In the RASA framework, the first step is converting the action logs to events. Algorithm 1 presents the generation of event list dictionary which stores the status transitions of all factors. As discussed in Section 3.2, an action log contains a factor (e.g., a person, document, or device), an action (e.g., enter, read, or exit), and a location. For each action log, the algorithm determines the previous event, creates a new event according to whether the action is to move the factor in or out, and finally updates the statuses of all elements of the new event. It is worth to note that factors do not include locations as opposed to the element of an event. The event list dictionary is further processed by Algorithm 2 to generate $Freq_{a_i,b_j}$ which counts the number of times that $a_i$ and $b_j$ coexist in the same event and $Dur_{a_i,b_j}$ which records the duration that two elements occur in the same event.

$$C^{Freq}_{a_i,b_j} = \frac{Freq_{a_i,b_j}}{\max_i Freq_{a_i,b_j}} \tag{1}$$

$$C^{Dur}_{a_i,b_j} = \frac{Dur_{a_i,b_j}}{\max_i Dur_{a_i,b_j}} \tag{2}$$

---

**Algorithm 1:** Event List Dictionary Generation

---

**Input:** *actionLogs*
**Output:** *eventListDict*
// Create a list if visit missing keys
*eventListDict* ← *EmptyDict*;
**foreach** *actionLog in actionLogs* **do**
  *event* ← *eventListDict*[*actionLog.loc*].*lastElement*;
  **if** *event not found* **then**
    | *event* ← {*actionLog.loc*}
  **end**
  **if** *action is moving a factor in (e.g., enter)* **then**
    | *event* ← *event* ∪ {*actionLog.factor*};
  **end**
  **else** *action is moving a factor out (e.g., exit)*
    | *event* ← *event* − {*actionLog.factor*}
  **end**
  **foreach** *elem in event* **do**
    | *eventListDict*[*elem*].*add*(*event*);
  **end**
**end**
**return** *eventListDict*

---

If $Freq_{a_i,b_j}$ and $Dur_{a_i,b_j}$ are considered, they can be abstracted as $G_{a_i,b_j}$. Then, Eq.3 represents a coupling matrix (not yet normalized). The time window of the distribution is determined by the tuple ($k_{start}$, $k_{end}$). $T_{E_k}$ indicates the duration of each event. Note that $A$ and $B$ should satisfy $Var(A) > Var(B)$. The rationale behind this condition is as follows: according to Eq. 1 and Eq.2, these two formulas asymmetrically normalize the values. That said, in order to make coupling values comparable, the differences between the elements which have larger variance need to be reduced.

---

**Algorithm 2:** $Freq_{a_i,b_j}$ and $Dur_{a_i,b_j}$ Generation

---

**Input:** $eventListDict$, $a_i$, $b_j$
**Output:** $Freq_{a_i,b_j}$ and $Dur_{a_i,b_j}$
$eventList \leftarrow eventListDict[a_i]$;
$Freq_{a_i,b_j} \leftarrow 0$;
$Dur_{a_i,b_j} \leftarrow 0$;
$coupledFlag \leftarrow false$;
$startTime \leftarrow 0$;
**foreach** $event$ $in$ $eventList$ **do**
    **if** $b_i \in event$ $and$ $coupledFlag = false$ **then**
        $coupledFlag \leftarrow true$;
        $Freq_{a_i,b_j} \leftarrow Freq_{a_i,b_j} + 1$;
        $startTime \leftarrow event.time$
    **end**
    **if** $b_i not in event$ $and$ $coupledFlag = true$ **then**
        $coupledFlag \leftarrow false$;
        $Dur_{a_i,b_j} \leftarrow event.time - startTime$;
    **end**
**end**
**return** $Freq_{a_i,b_j}, Dur_{a_i,b_j}$

---

$$\begin{bmatrix} G_{a_1,b_1} & G_{a_1,b_2} & ... & G_{a_1,b_{N_B}} \\ G_{a_2,b_1} & G_{a_2,b_2} & & G_{a_2,b_{N_B}} \\ ... & ... & ... & ... \\ G_{a_{N_A},b_1} & G_{a_{N_A},b_2} & ... & G_{a_{N_A},b_{N_B}} \end{bmatrix} \qquad (3)$$

$$\begin{bmatrix} C_{a_1,b_1} & C_{a_1,b_2} & ... & C_{a_1,b_{N_B}} \\ C_{a_2,b_1} & C_{a_2,b_2} & & C_{a_2,b_{N_B}} \\ ... & ... & ... & ... \\ C_{a_{N_A},b_1} & C_{a_{N_A},b_2} & ... & C_{a_{N_A},b_{N_B}} \end{bmatrix} \qquad (4)$$

Since the action logs contain four factors–namely people, document, device and location–theoretically $\binom{4}{2} = 6$ different couplings can be obtained. These are: $C_{device,location}$, $C_{device,document}$, $C_{document,location}$, $C_{people,location}$, $C_{people,device}$, and $C_{people,document}$. As mentioned earlier, coupling values are asymmetric. For instance $C_{A,B}$ is applied instead of $C_{B,A}$ in case the variance of the elements over $B$ is greater than those in $A$.

In case an object, such as a physician, interacts/couples with other objects or locations equally, its couplings will tend towards 1. If two objects less frequently interact, their coupling tends towards 0, which means the risk associated to their coexistence is high.

Fig. 5 illustrates how the coupling distribution in a matrix can be skewed.

When multiple coupling values are of the same type, lower coupling value is used to ensure that the impact of risky elements will not be alleviated by safe ones. This is formulated in Eq. 5. For instance, given $C_{a_i,b_k} < C_{a_j,b_k}$, when $a_i$ and $a_j$ appear with $b_k$ at the same time, $F_n^{A,B} = C_{a_i,b_k}$. Specifically, if $E_i \subseteq E_j$, then $Risk(E_i) \leq Risk(E_j)$.

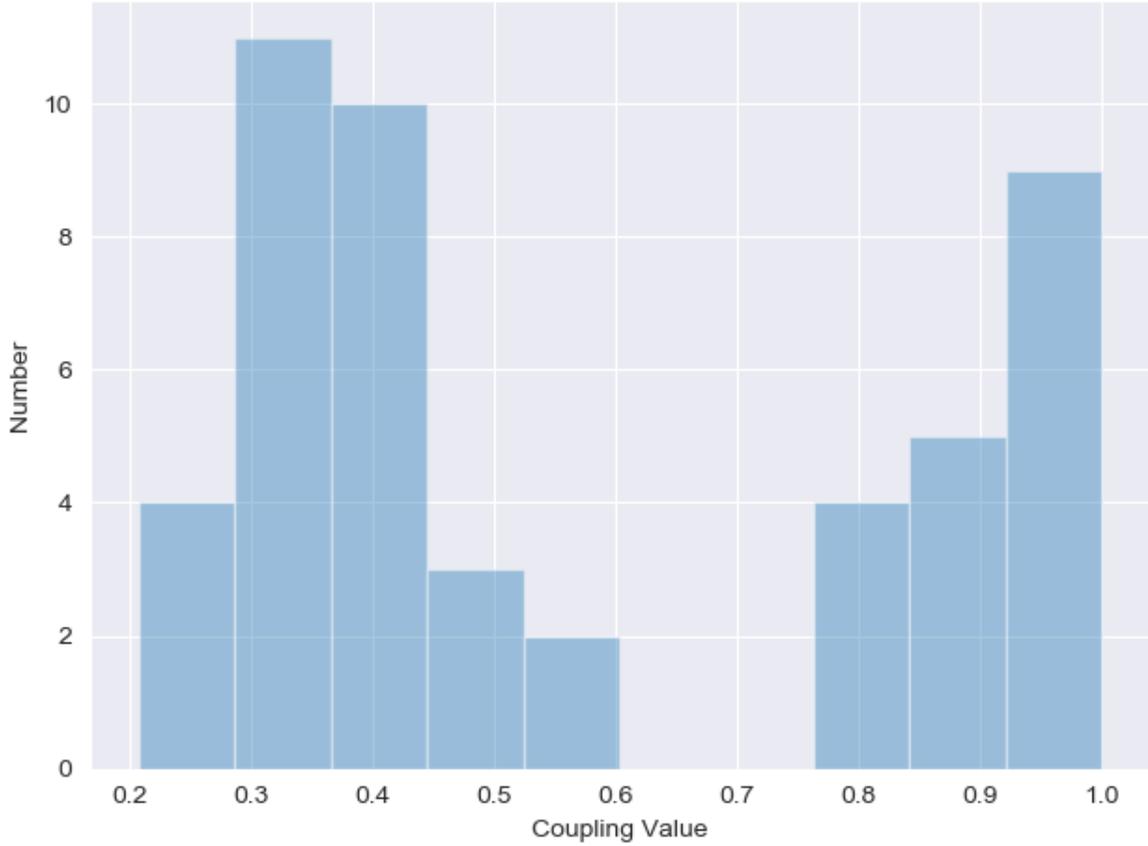

Fig. 5. An Example of Distribution of Coupling Matrix

For instance, as shown in Fig. 6, $p_1$ and $p_3$ co-exist with $doc_1$ in $E_{13}$. $C_{p_1,doc_1}$ and $C_{p_3,doc_1}$ are determined according to historical records and $C_{p_1,doc_1} > C_{p_3,doc_1}$. Therefore, for this event, $F_{13}^{PpI,Doc} = C_{p_3,doc_1}$

$$F_n^{A,B} = \min_{i,j} C_{a_i,b_j}, \forall a_i, b_j \in E_n \tag{5}$$

Our final goal is to label each cluster of couplings with a risk level, which will be described in the next section. To be able to associate clusters with the risk, we need to first assign risk levels to each individual coupling. Therefore, for each coupling matrix, we map the coupling values to onto three risk levels as high, medium or low risk. To achieve this, we set a threshold as Eq. 6 and apply mean value binning over the coupling values.

$$Threshold_{High} = mean - \alpha * stdev \tag{6}$$

To formulate the risk level of each event, several strategies are applied, one of which is to code high as three, medium as two and low as one. This is followed by calculating the average risk of each event. Then the average risk will be marked as high, medium, low-risk. The other one is to directly code the risk level of each feature, which can directly represent features' pattern. The latter one is used to evaluate the results. Even if the risk level of each coupling value is inferred, still, it can not be used as ground truth because we introduce human subjectivity and intuition into this inference.

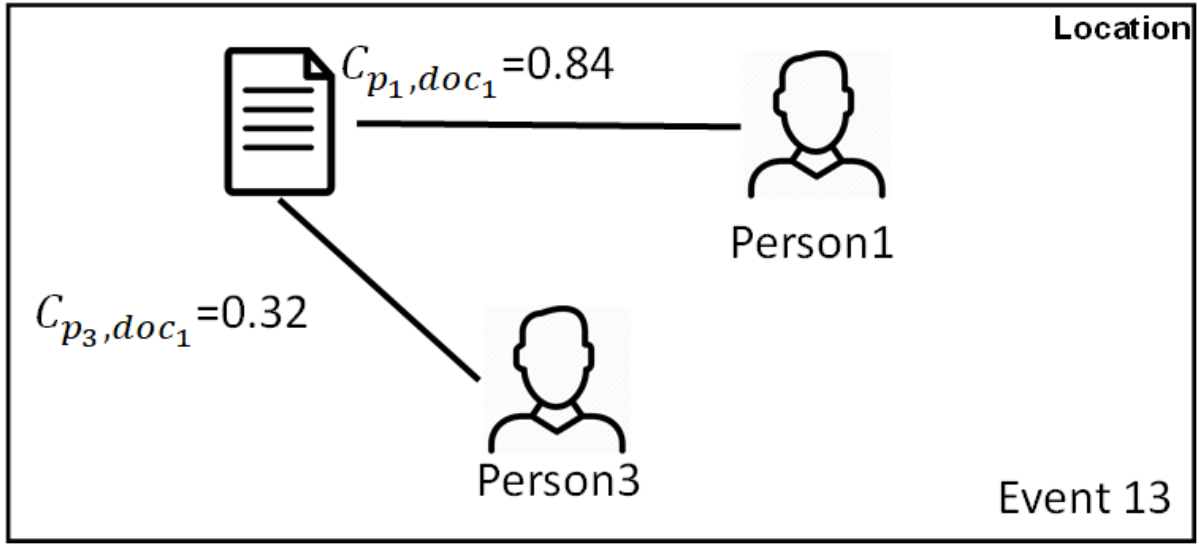

Fig. 6. Coupling Feature Example

Table 4. Map Cluster Risk Values to Risk Levels

| CRV | 1 | 1~1.5 | 1.5~2 | 2 | 2~2.5 | 2.5~3 | 3 |
|-----|---|-------|-------|---|-------|-------|---|
| RL | L | LM | ML | M | MH | HM | H |

Conventionally, researchers may adopt a risk matrix [15] to evaluate the risk level, e.g. *Risk ∝ Probability ∗ Impact*. However, this study infers risk by defining association between various objects and situations in the action logs. Thus, the use of conventional risk formulation is left to future work along with further investigations on the quantification of the impact.

### 3.4 Clustering of Couplings and Labeling of the Clusters

RASA framework clusters the couplings into groups and uses the cluster members' risk levels to determine the label of each cluster. Note that non-clustered couplings are less frequently observed, and so possess higher risk.

Most clustering algorithms put similar distance samples into one cluster. In this work, risk levels of couplings are used as the distance metric in cluster formation.

The cluster risk value (CRV) is determined based on the following formula:

$$CRV = \frac{N_H * C_H + N_M * C_M + N_L * C_L}{N_H + N_M + N_L} \quad (7)$$

where $N_H$, $N_M$, $N_L$ denote the number of high, medium and low-risk features, respectively; and $C_H$, $C_M$, $C_L$ are the codes that are assigned to high, medium and low-risk levels. $C_H$, $C_M$, $C_L$ corresponds to 3, 2, 1, respectively, 3 denoting a higher risk. This is followed by mapping these values to Low (L), Low-Medium (LM), Medium-Low (ML), Medium (M), Medium-High (MH), High-Medium (HM), and High (H) risk levels according to Table 4. In case a cluster is labeled high-risk, all actions in this cluster would be denied. Otherwise, the clusters with medium or low-risk labels contain the actions that may be permitted. The decision for medium risk clusters depends on the coupling feature-based risk levels for each individual point.

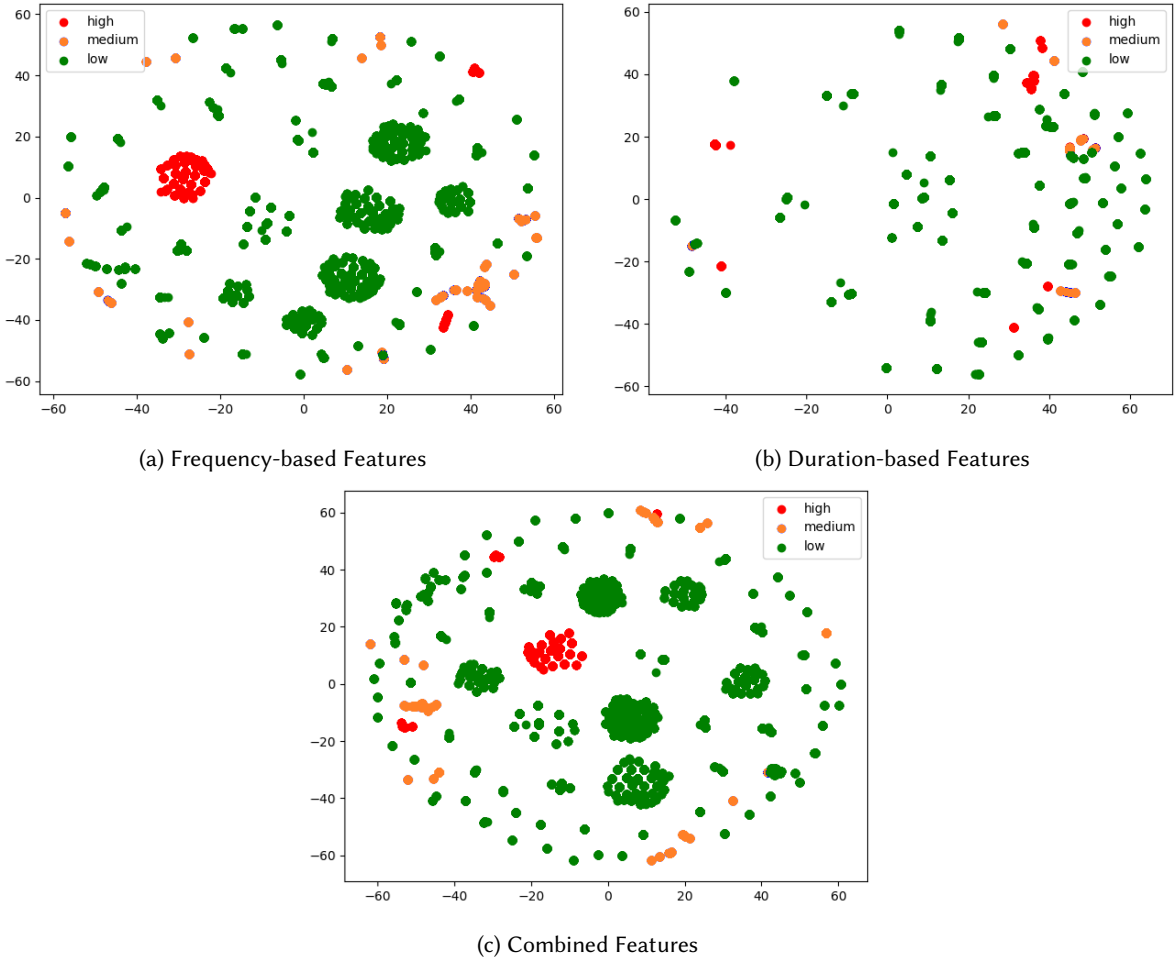

(a) Frequency-based Features

(b) Duration-based Features

(c) Combined Features

Fig. 7. Features Visualization using t-SNE

## 4 NUMERICAL RESULTS

In this section we evaluate the initial promise of the RASA approach by comparing the decisions against an heuristic policy, and also validated with two supervised learning methods, namely decision tree and Support Vector Machine (SVM).

The action logs used in this work contain one device; hence the available features are Location-People, Location-Device, Location-Document, and Document-People. Because the action logs cover a single device, only $C_{document,location}$, $C_{people,location}$, $C_{people,people}$, and $C_{people,document}$ are used in the evaluations. We present the results under using duration-based features, frequency-based features and the combined (i.e. frequency and duration-based) features. In addition, three different clustering algorithms are applied, which are Gaussian mixture models (GMM) [36], Agglomerative Hierarchical [43], and DBSCAN [13]. The target is to cluster these features (couplings) and label the clusters with risk levels. The aim of the clustering algorithm is to form the least possible number of clusters where within each cluster co-existence of high and low risk patterns is avoided.

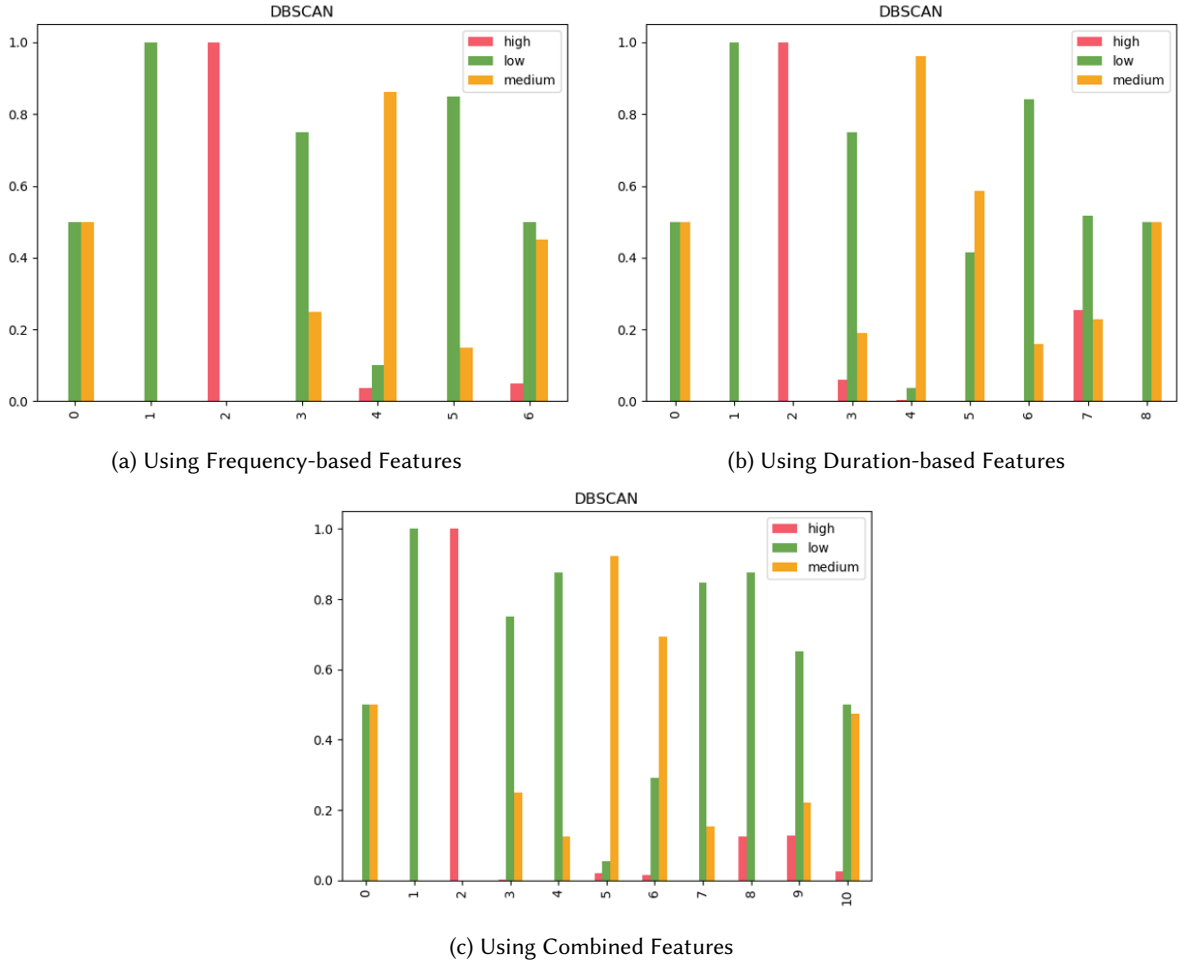

(a) Using Frequency-based Features

(b) Using Duration-based Features

(c) Using Combined Features

Fig. 8. DBSCAN Result of Feature-by-Feature Risk Level using Different Features

## 4.1 Numerical Results of Clustering and Labeling

This subsection presents visualization of features followed by feature-by-feature risk of each cluster in the action logs. Then, results under different clustering algorithms are discussed, and finally cluster risk levels are presented

*4.1.1 Visualization of features.* Fig. 7 illustrates the features visualized via using t-SNE [41][34]; features are presented by color-codes to show risk levels of samples (red=high-risk, green=low-risk and orange=medium-risk).

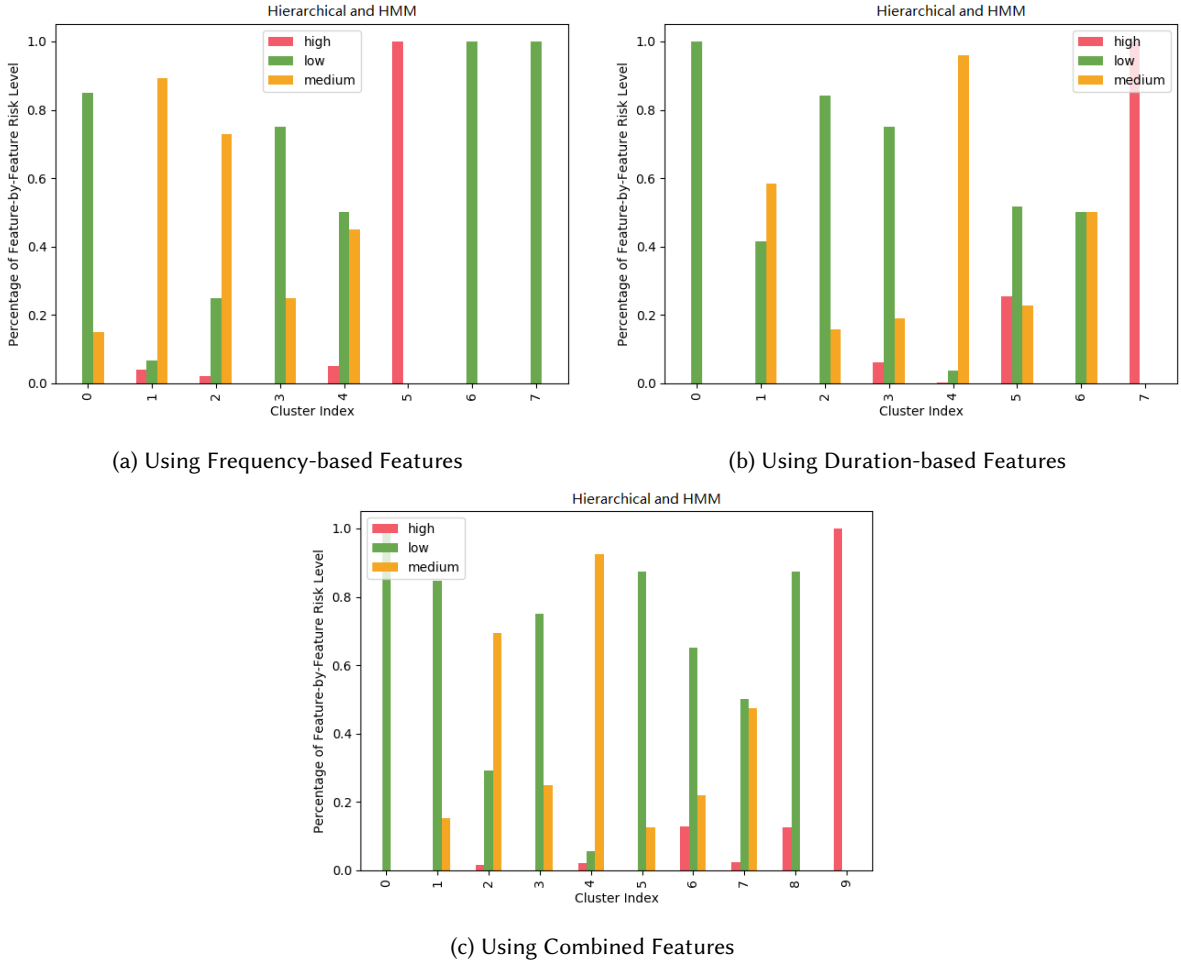

(a) Using Frequency-based Features

(b) Using Duration-based Features

(c) Using Combined Features

Fig. 9. Hierarchical and GMM clustering results of feature-by-feature risk level using different features

In Fig. 7 (a–c), most of the data points are of low risk level (green) whereas the medium and high average-risk samples are less frequently observed. t-SNE maps every multi-dimensional input in a data set onto two or three dimensions. When t-SNE-based modeling is complete, similar data are mapped close to each other whereas dissimilar data are mapped onto distant points [29].

*4.1.2 Feature-by-Feature risk of clusters.* We calculate and present the percentage of feature-by-feature risk level (high/medium/low risk) found at each cluster when we use DBSCAN, Hierarchical and GMM clustering. Note that hierarchical and GMM clustering schemes generate the same results, as shown in Fig. 9. It is not viable to apply the frequency or duration as the only metric to evaluate objects' relationship and their risk levels. In this study, we also consider the circumstances under which people may stay in a room for a long time while they enter and exit other rooms with high frequency. Therefore, the frequency-based features are also expected to affect the risk assessment. Since the duration-based coupling features and frequency-based features show distinctive perspective of the action logs, this study also employs the combined features defined as concatenation of two features. Clustering results, and risk value results are generated by using three different features. Fig. 8 and Fig. 9 present the percentage of feature-by-feature risk levels of clusters by using frequency-based, duration-based, and combined features. When Fig. 8 and Fig. 9 are compared when given the same number of clusters, DBSCAN, Hierarchical and HMM provide similar results, although DBSCAN results suggest it may increased noise.

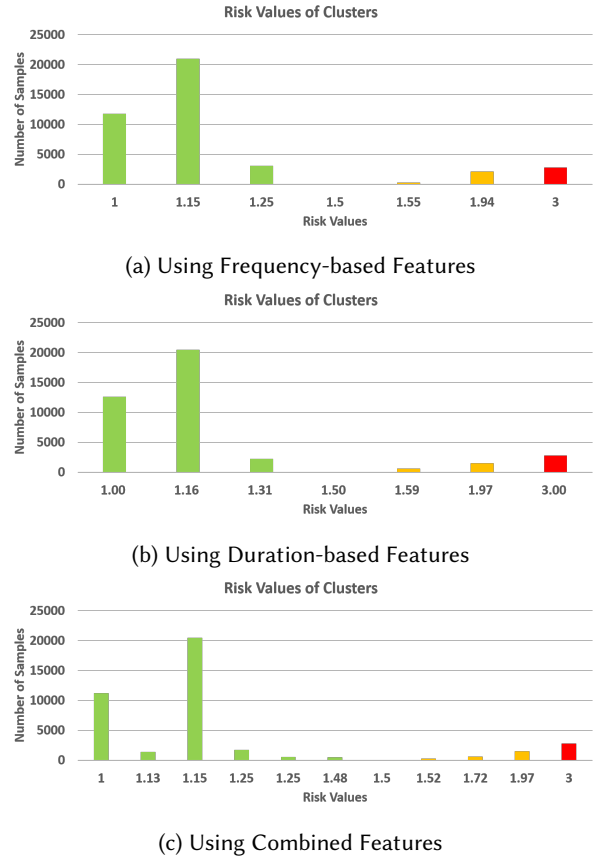

(a) Using Frequency-based Features

(b) Using Duration-based Features

(c) Using Combined Features

Fig. 10. DBSCAN Cluster Risk Values using Different Features

Both Fig. 8 and Fig. 9 illustrate the diversity of three features; however the results of these three features have similar pattern. Low-risk events are dominant, other types of risk are minor, which is what one would hope is a realistic distribution of event risks in normal clinical practice. This pattern also results in the similarity of the three figures in Fig. 10. It is worth to note that the occurrence of this pattern is only due to the simulated action logs used in this study. When different action logs are employed, the result of three features may lead to a more significant variation.

*4.1.3 Clustering algorithms: (HMM and DBSCAN).* The strength of DBSCAN is that we do not need to specify the number of clusters upfront while in hierarchical and GMM number of desired clusters are given. If there were more factors in the action logs, the number of features would increase, hence labeling of action clusters by using DBSCAN is generalizable. Indeed, outliers are expected, however, the outliers are not treated as high-risk samples directly, because the events maybe the sub-events of another cluster. For instance, in Tables 5,6 and 7, the sample in the cluster -1, does not have documents in this event; hence there is no possibility of leaking sensitive data in such an event. Nevertheless, in practice we expect that, as a matter of policy, outliers would be escalated to the security operations and policy experts.

*4.1.4 Cluster Risk Levels.* As clusters and their feature-by-feature risk levels are obtained, risk values of clusters are calculated according to Eq. 7 and binned into 7 groups: Low, Low-medium, medium-low, medium, medium-high, high-medium, and high. Since the actions in medium and low-risk level clusters will be permitted, we intend to prevent high average-risk (defined in Section 3.4) samples occur in medium and low-risk level clusters. Although the average-risk level is not a kind of ground truth, if it shows high-risk, then most of the features of the sample raise alert or medium-risk. As explained in the feature extraction section, the coupling value of medium-risk is similar to that of high-risk. There is no obvious boundary between these two risk levels, and clustering algorithms provide some potential risk patterns in the action logs.

The risk values and levels of each cluster are shown on in Tables 5, 6, and 7. The cluster risk values are sorted and plotted as illustrated in Fig. 10. This paper further compares the outcomes of three features in Fig. 11. Most of them are safe clusters whereas only one cluster raises a high-risk alert. Based on this, the risk value of the entire dataset (i.e. action log) can be calculated as well, which is 1.28 in this case.

Table 5. DBSCAN Result using Duration-based Features

| Index | Risk Value | Risk Level | Number of samples |
|-------|------------|------------|-------------------|
| -1 | 1.5 | LM | 1 |
| 0 | 1 | L | 12612 |
| 1 | 3 | H | 2804 |
| 2 | 1.31 | LM | 2234 |
| 3 | 1.97 | ML | 1492 |
| 4 | 1.59 | ML | 608 |
| 5 | 1.16 | LM | 20489 |
| 6 | 1.74 | ML | 480 |
| 7 | 1.5 | LM | 246 |

Table 6. DBSCAN Result using Frequency-based Features

| Index | Risk Value | Risk Level | Number of samples |
|-------|------------|------------|-------------------|
| -1 | 1.5 | LM | 1 |
| 0 | 1 | L | 11754 |
| 1 | 3 | H | 2804 |
| 2 | 1.25 | LM | 3082 |
| 3 | 1.94 | ML | 2100 |
| 4 | 1.15 | LM | 20979 |
| 5 | 1.55 | ML | 246 |

Table 7. DBSCAN Result using Combined-based Features

| Index | Risk Value | Risk Level | Number of samples |
|-------|-----------|-----------|-------------------|
| -1 | 1.5 | LM | 1 |
| 0 | 1 | L | 11222 |
| 1 | 3 | H | 2804 |
| 2 | 1.25 | LM | 1702 |
| 3 | 1.13 | LM | 1390 |
| 4 | 1.97 | ML | 1492 |
| 5 | 1.72 | ML | 608 |
| 6 | 1.15 | LM | 20489 |
| 7 | 1.25 | LM | 532 |
| 8 | 1.48 | LM | 490 |
| 9 | 1.52 | ML | 246 |

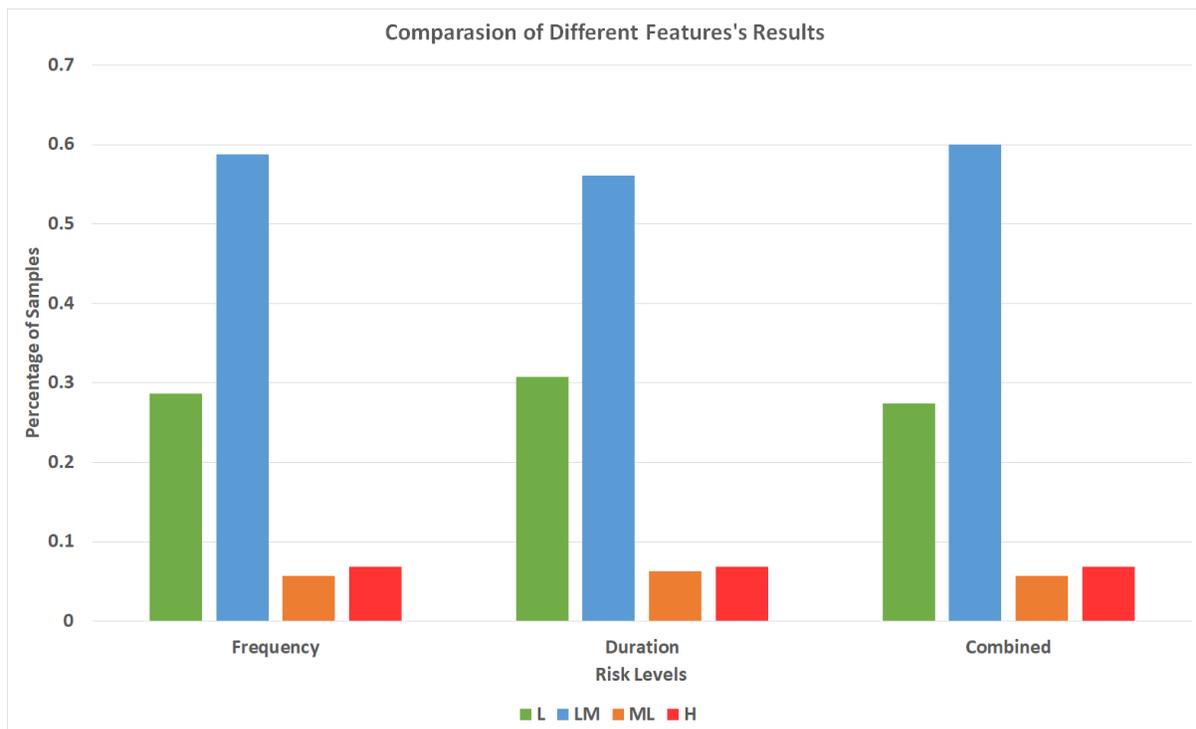

Fig. 11. Comparison of Different Results

## 4.2 Validation via Supervised Learning

In the final step, as shown in the Fig. 13, this work utilizes supervised learning methods, decision tree and SVM, to validate the results of unsupervised learning and to verify whether the coupling idea can be generalized.

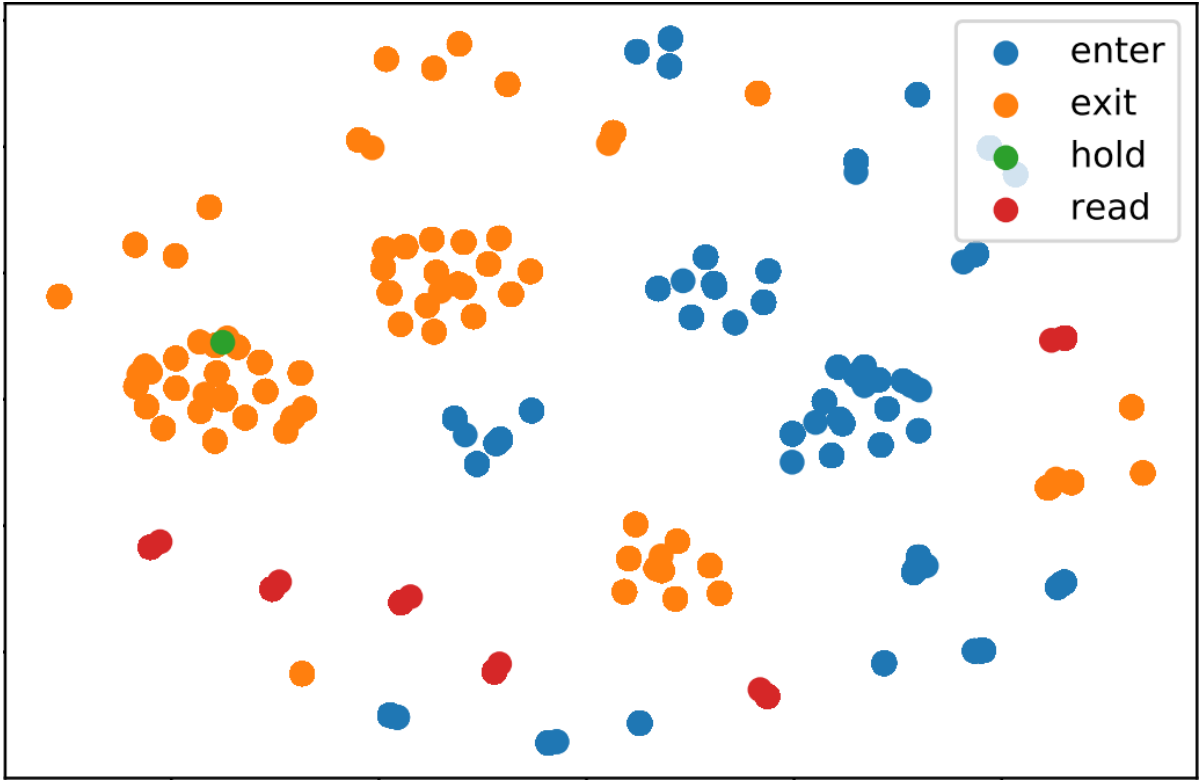

Fig. 12. Visualization of Raw Actions in Dataset 2

As shown in Fig. 14, when the model is trained with dataset 1, and tested with dataset 2, 100% training accuracy and 99.86% test accuracy are achieved. As Fig. 3 and Fig. 12 present, despite the distinct distributions of dataset 1 and 2, the proposed methodology can still function well.

Support Vector Machine (SVM) is also employed and it achieves 99.95% test accuracy.

## 4.3 Access Control Policy-based Decisions

The performance of RASA framework is validated against a policy-based access control scheme. Policy-based decisions can be considered as a Attribute-Based Access Control (ABAC) [20] approach, utilizing the four frequency-based coupling features introduced above in addition to the traffic, which indicates the number of people in the current event, and co-existence, which is triple-coupling, people-document-location coupling. Eventhough the traffic does not entail any coupling, it is possible to use Eq. 6, and the average value to determine the risk level of each coupling value.

Other features would further transform into risk as well. The policy maps the features' risk onto three risk categories: Device Risk, Environment Risk, and Action Risk. Furthermore it applies a weighted sum of these risks together to determine the overall risk.

$$R_{Dev} = 0.5 * R_{dev,loc}^{Freq}$$

$$R_{Env} = 0.5 * R_{traffic} + 0.5 * R_{co-existence}^{Freq}$$

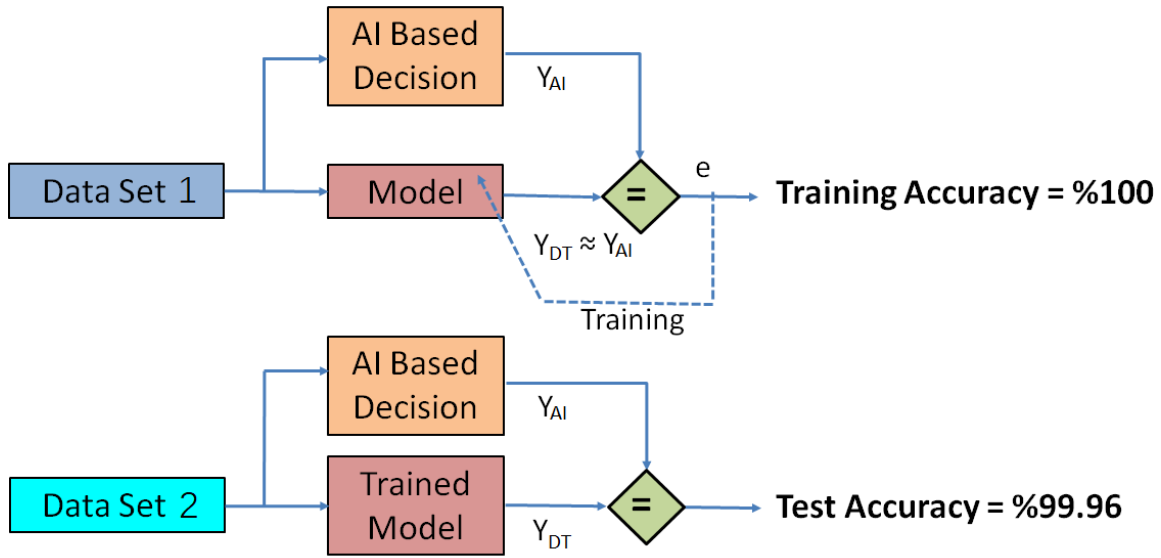

Fig. 13.  Supervised Learning Validation

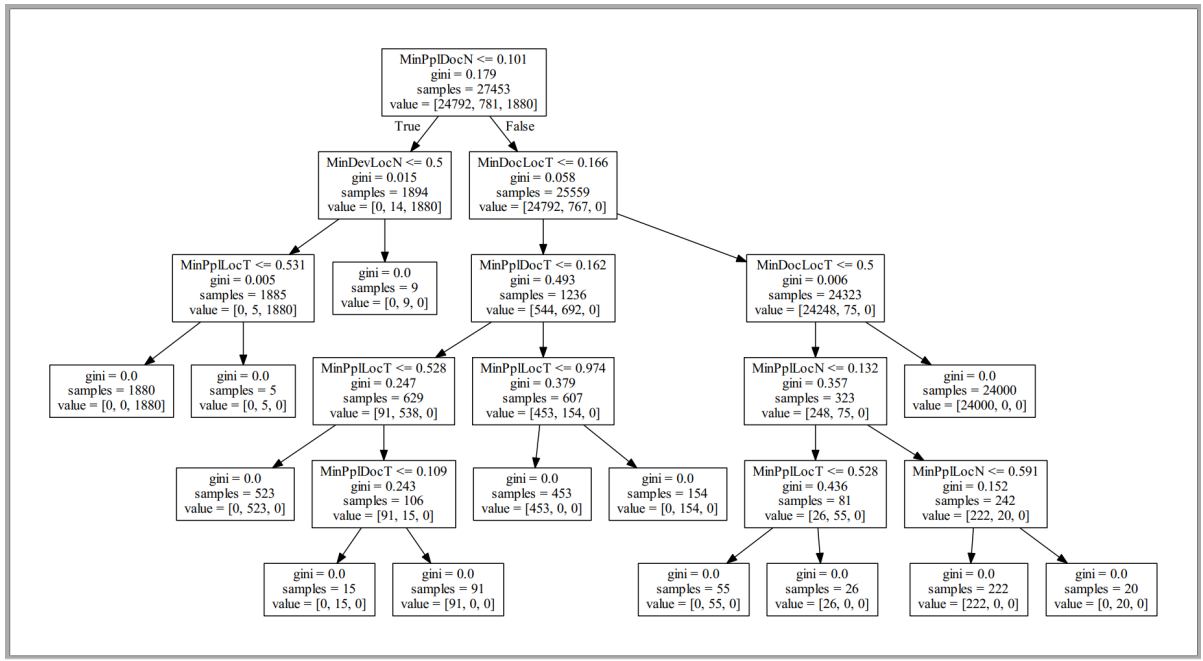

Fig. 14.  Trained Decision Tree

Table 8. Comparison between Policy and DBSCAN result using Frequency-based Features

| Policy | Permit | Deny | |
|---|---|---|---|
| DBSCAN | Medium-Low | Medium-Low | High |
| Numer | 586 | 23 | 2800 |
| Consistency | 100% | 99.19% | |
| Overall Consistency | 99.32% | | |

Table 9. Comparison between Policy and DBSCAN result using Duration-based Features

| Policy | Permit | Deny | |
|---|---|---|---|
| DBSCAN | Medium-Low | Medium-Low | High |
| Numer | 586 | 25 | 2798 |
| Consistency | 100% | 99.19% | |
| Overall Consistency | 99.27% | | |

Table 10. Comparison between Policy and DBSCAN result using Combined Features

| Policy | Permit | Deny | |
|---|---|---|---|
| DBSCAN | Medium-Low | Medium-Low | High |
| Numer | 586 | 23 | 2800 |
| Consistency | 100% | 99.19% | |
| Overall Consistency | 99.32% | | |

$$R_{Act} = 0.5 * R_{doc,loc}^{Freq} + 0.5 * R_{doc,time}$$

$$R_{Overall} = 0.3 * R_{Dev} + 0.4 * R_{Env} + 0.3 * R_{Act}$$

Due to the lack of labels or other methods to prove the efficiency of these methods, to examine the performance of the proposed scheme, the unsupervised method and policy-based method are also compared with each other[26]. By tuning the parameters of policy, we also investigated the outcomes of RASA and policy and analyzed the action log. As shown in the Table. 8, Table. 9, and Table. 10, the frequency-based, duration-based, and combined coupling features show 99.32%, 99.27%, and 99.32% overall consistency, respectively. Since the policy uses the triple factor coupling, it is more likely to detect the rare cases but meanwhile, it is the only triple factor coupling since only one device exists in the action log. When the number of coupling factors increases, the types of couplings will reduce, and such a situation would occur where the frequency or duration is not enough to make statistical decision because the frequency or duration disperse on more coupling combinations.

## 5 CONCLUSION AND FUTURE WORK

In this paper, we propose the Risk-Aware Smart Access Control (RASA) framework, which defines a simple overall structure to learn use contexts from observed resource access within cyberphysical contexts, and use simple parameters to define a context-dependent and risk-based authorization scheme based on those contexts. The basic framework can be utilized to extend other access control regimes that lack sufficient support for security policy analysts to craft and maintain context-dependent authorization policies. RASA utilizes combined cyber and physical activity logs, converts action sequence to events which contain contextual information, and identifies common relations through a coupling approach based on duration and frequency of interactions.

Every coupling is assigned a risk value which results from the intuition that frequently occurring interactions can be considered as low-risk and rarely occurring actions as high-risk. RASA further feeds these couplings to clustering algorithms, after which the level of risk for a cluster is determined based on the risk levels of cluster members (i.e. couplings). Thereafter RASA denies actions in high-risk clusters and permits safe actions. One of the insights from our study in a simulated clinical environment is the importance of the selection of couplings on risk inference.

Three different features are explored in this study, namely, frequency-based, duration-based and the combination of frequency and duration. According to clustering and labelling results, these features can exhibit distinct perspectives of risk. We show that combined features can more comprehensively represent the risk of events and actions.

It may be helpful in future work to define other risk factor compositions and understand better their interactions.

RASA is purposefully designed to require little in the way of security policy analyst input–for example, no extensive labeling of actions into risk is required to develop a training set. Our evaluation established a basic validation of the promise of building on this approach.

In the tuning process, rule-based policy results are compared with the outcomes of RASA, and rule-based policy and RASA are updated in each iteration. After multiple tuning iterations, the decisions of two approaches achieved > 99% consistency. Furthermore, to validate the results of RASA, when two supervised methods (Decision tree and SVM) are trained and tested by using different action logs labeled by RASA, 99.95% test accuracy is achieved.

Future work could seek evaluation against data derived from authentic use cases. Similar approaches to automatic inference of authorization contexts and associated risks may be particularly applicable in rich cyber-physical environments where sensors are deployed to generate activity logs by monitoring physical actions, in addition to common network and endpoint cyber action logs. Such sensor-rich environments are becoming increasingly common in extremely diverse cyberphysical contexts, including clinical, industrial, home, automotive, smart-building, and smart-city contexts. As these cyberphysical systems grow in size and complexity, human-written context-dependent authorization access rules will not scale, nor will expert risk analysis be cost effective and reliable without evidence-based risk assessment of such accesses. The RASA framework proposed in this paper may be a stepping stone to rich exploration of similar approaches to address such growing problems in cyberphysical security.

## ACKNOWLEDGMENT

This study has been supported in part by the Natural Sciences and Engineering Research Council (NSERC) of Canada under the DISCOVERY program, and Ontario Centres of Excellence Voucher for Innovation Productivity 1 program. The authors would like to thank Yueqian Zhang for collaborating with the team to provide the security policy updates.

## APPENDIX

In this section, we present further detailed results about coupling matrices and normalized coupling matrices for each pair of coupled factors.

By following Algorithm 1 and Algorithm 2, the action logs are transformed into events, and the frequency and duration information are extracted. As shown in Table 11 and Table 12, we present $Freq_{a_i,b_j}$ and $Dur_{a_i,b_j}$ in $Freq_{A,B}$ and $Dur_{A,B}$ coupling matrices, respectively. By applying Eq. 1 and Eq. 2, the coupling matrices are transformed into normalized coupling matrices listed in Table 13 and Table 14. Since one device exists in the dataset, four couplings are selected as features (i.e., $C_{Doc,Loc}$, $C_{Ppl,Doc}$, $C_{Ppl,Loc}$ and $C_{Ppl,Ppl}$). By viewing both the frequency-based and duration-based coupling matrices, one can have a clear understanding of how factors interact with each other.

Table 11. Frequency-based Coupling Matrices

(a) $Freq_{Dev,Loc}$

|  | Device1 |
|---|---|
| Location1 | 5065 |
| Location2 | 0 |
| Location3 | 0 |
| Location4 | 854 |
| Location5 | 0 |
| Location6 | 0 |
| Location7 | 0 |

(b) $Freq_{Doc,Loc}$

|  | Doc1 | Doc2 | Doc3 | Doc4 | Doc5 | Doc6 |
|---|---|---|---|---|---|---|
| Location1 | 782 | 846 | 957 | 834 | 794 | 852 |
| Location2 | 0 | 0 | 0 | 0 | 0 | 0 |
| Location3 | 0 | 0 | 0 | 0 | 0 | 0 |
| Location4 | 215 | 232 | 233 | 298 | 245 | 241 |
| Location5 | 0 | 0 | 0 | 0 | 0 | 0 |
| Location6 | 0 | 0 | 0 | 0 | 0 | 0 |
| Location7 | 0 | 0 | 0 | 0 | 0 | 0 |

(c) $Freq_{Ppl,Doc}$

|  | Ppl1 | Ppl2 | Ppl3 | Ppl4 | Ppl5 | Ppl6 | Ppl7 | Ppl7 |
|---|---|---|---|---|---|---|---|---|
| Doc1 | 997 | 1543 | 45 | 38 | 116 | 47 | 63 | 46 |
| Doc2 | 1078 | 1647 | 55 | 24 | 51 | 40 | 55 | 116 |
| Doc3 | 1190 | 1937 | 70 | 116 | 55 | 42 | 63 | 37 |
| Doc4 | 1132 | 1601 | 79 | 45 | 49 | 74 | 203 | 46 |
| Doc5 | 1039 | 1560 | 140 | 37 | 46 | 41 | 70 | 62 |
| Doc6 | 1093 | 1712 | 52 | 35 | 35 | 144 | 75 | 48 |

(d) $Freq_{Ppl,Loc}$

|  | Ppl1 | Ppl2 | Ppl3 | Ppl4 | Ppl5 | Ppl6 | Ppl7 | Ppl8 |
|---|---|---|---|---|---|---|---|---|
| Loc1 | 5066 | 4936 | 159 | 167 | 201 | 166 | 161 | 168 |
| Loc2 | 0 | 0 | 1 | 0 | 0 | 0 | 0 | 0 |
| Loc3 | 0 | 0 | 0 | 1 | 0 | 0 | 0 | 0 |
| Loc4 | 854 | 818 | 25 | 22 | 28 | 26 | 36 | 25 |
| Loc5 | 0 | 0 | 0 | 0 | 0 | 1 | 0 | 0 |
| Loc6 | 0 | 0 | 0 | 0 | 0 | 0 | 1 | 0 |
| Loc7 | 0 | 0 | 0 | 0 | 0 | 0 | 0 | 1 |

Table 11. Frequency-based Coupling Matrices

(e) $Freq_{Ppl,Ppl}$

|        | Ppl1  | Ppl2  | Ppl3 | Ppl4 | Ppl5 | Ppl6 | Ppl7 | Ppl8 |
|--------|-------|-------|------|------|------|------|------|------|
| Ppl1   |       | 10001 | 259  | 190  | 222  | 239  | 304  | 218  |
| Ppl2   | 10001 |       | 243  | 171  | 231  | 210  | 284  | 207  |
| Ppl3   | 259   | 243   |      | 11   | 6    | 9    | 15   | 9    |
| Ppl4   | 190   | 171   | 11   |      | 6    | 6    | 11   | 4    |
| Ppl5   | 222   | 231   | 6    | 6    |      | 7    | 18   | 11   |
| Ppl6   | 239   | 210   | 9    | 6    | 7    |      | 17   | 7    |
| Ppl7   | 304   | 284   | 15   | 11   | 18   | 17   |      | 15   |
| Ppl8   | 218   | 207   | 9    | 4    | 11   | 7    | 15   |      |

Table 12. Duration-based Coupling Matrices

(a) $Dur_{Dev,Loc}$

|           | Device1           |
|-----------|-------------------|
| Location1 | 335384.62449996645 |
| Location2 | 0.0               |
| Location3 | 0.0               |
| Location4 | 56341.1491895333  |
| Location5 | 0.0               |
| Location6 | 0.0               |
| Location7 | 0.0               |

(b) $Dur_{Doc,Loc}$

|      | Doc1  | Doc2  | Doc3  | Doc4  | Doc5  | Doc6  |
|------|-------|-------|-------|-------|-------|-------|
| Loc1 | 51227 | 54270 | 66317 | 52308 | 52734 | 58526 |
| Loc2 | 0.0   | 0.0   | 0.0   | 0.0   | 0.0   | 0.0   |
| Loc3 | 0.0   | 0.0   | 0.0   | 0.0   | 0.0   | 0.0   |
| Loc4 | 7827  | 9155  | 9150  | 11185 | 9754  | 9237  |
| Loc5 | 0.0   | 0.0   | 0.0   | 0.0   | 0.0   | 0.0   |
| Loc6 | 0.0   | 0.0   | 0.0   | 0.0   | 0.0   | 0.0   |
| Loc7 | 0.0   | 0.0   | 0.0   | 0.0   | 0.0   | 0.0   |

(c) $Dur_{Ppl,Doc}$

|      | Ppl1  | Ppl2 | Ppl3 | Ppl4 | Ppl5 | Ppl6 | Ppl7 | Ppl8 |
|------|-------|------|------|------|------|------|------|------|
| Doc1 | 59054 | 1544 | 1344 | 1134 | 3840 | 1468 | 1887 | 1439 |
| Doc2 | 63425 | 1647 | 1711 | 720  | 1593 | 1202 | 1713 | 3781 |
| Doc3 | 75468 | 1935 | 2285 | 3958 | 1649 | 1262 | 1954 | 1171 |
| Doc4 | 63494 | 1600 | 2486 | 1354 | 1589 | 2214 | 7106 | 1556 |
| Doc5 | 62489 | 1560 | 4682 | 1171 | 1379 | 1290 | 2096 | 1857 |
| Doc6 | 67763 | 1712 | 1616 | 1047 | 1111 | 4560 | 2311 | 1502 |

Table 12. Duration-based Coupling Matrices

(d) $Dur_{Ppl,Loc}$

|      | Ppl1   | Ppl2   | Ppl3   | Ppl4  | Ppl5   | Ppl6   | Ppl7   | Ppl8   |
|------|--------|--------|--------|-------|--------|--------|--------|--------|
| Loc1 | 335385 | 345851 | 4772   | 5002  | 6028   | 4978   | 4828   | 5042   |
| Loc2 | 0      | 0      | 3167   | 0     | 0      | 0      | 0      | 0      |
| Loc3 | 0      | 0      | 0      | 2801  | 0      | 0      | 0      | 0      |
| Loc4 | 56341  | 54442  | 121547 | 80092 | 107772 | 108641 | 159899 | 102243 |
| Loc5 | 0      | 0      | 0      | 0     | 0      | 4627   | 0      | 0      |
| Loc6 | 0      | 0      | 0      | 0     | 0      | 0      | 913    | 0      |
| Loc7 | 0      | 0      | 0      | 0     | 0      | 0      | 0      | 670    |

(e) $Dur_{Ppl,Ppl}$

|      | Ppl1  | Ppl2  | Ppl3  | Ppl4  | Ppl5  | Ppl6  | Ppl7  | Ppl8  |
|------|-------|-------|-------|-------|-------|-------|-------|-------|
| Ppl1 |       | 10001 | 14127 | 9386  | 11193 | 11998 | 17069 | 11309 |
| Ppl2 | 10001 |       | 12633 | 8251  | 11043 | 10840 | 15580 | 10140 |
| Ppl3 | 14127 | 12633 |       | 16476 | 9255  | 19966 | 37563 | 24455 |
| Ppl4 | 9386  | 8251  | 16476 |       | 16825 | 5055  | 19705 | 4385  |
| Ppl5 | 11193 | 11043 | 9255  | 16825 |       | 13669 | 23381 | 14461 |
| Ppl6 | 11998 | 10840 | 19966 | 5055  | 13669 |       | 34577 | 17834 |
| Ppl7 | 17069 | 15580 | 37563 | 19705 | 23381 | 34577 |       | 26891 |
| Ppl8 | 11309 | 10140 | 24455 | 4385  | 14461 | 17834 | 26891 |       |

Table 13. Normalized Frequency-based Coupling Matrices

(a) $C_{Dev,Loc}^{Freq}$

|           | Device1             |
|-----------|---------------------|
| Location1 | 1.0                 |
| Location2 | 0.0                 |
| Location3 | 0.0                 |
| Location4 | 0.1686080947680158  |
| Location5 | 0.0                 |
| Location6 | 0.0                 |
| Location7 | 0.0                 |

(b) $C_{Doc,Loc}^{Freq}$

|      | Doc1 | Doc2 | Doc3 | Doc4 | Doc5 | Doc6 |
|------|------|------|------|------|------|------|
| Loc1 | 1.0  | 1.0  | 1.0  | 1.0  | 1.0  | 1.0  |
| Loc2 | 0.0  | 0.0  | 0.0  | 0.0  | 0.0  | 0.0  |
| Loc3 | 0.0  | 0.0  | 0.0  | 0.0  | 0.0  | 0.0  |
| Loc4 | 0.27 | 0.27 | 0.24 | 0.36 | 0.31 | 0.28 |
| Loc5 | 0.0  | 0.0  | 0.0  | 0.0  | 0.0  | 0.0  |
| Loc6 | 0.0  | 0.0  | 0.0  | 0.0  | 0.0  | 0.0  |
| Loc7 | 0.0  | 0.0  | 0.0  | 0.0  | 0.0  | 0.0  |

Table 13. Normalized Frequency-based Coupling Matrices

(c) $C_{Ppl,Doc}^{Freq}$

|      | Ppl1 | Ppl2 | Ppl3 | Ppl14 | Ppl5 | Ppl6 | Ppl7 | Ppl8 |
|------|------|------|------|-------|------|------|------|------|
| Doc1 | 0.84 | 0.80 | 0.32 | 0.33  | 1.00 | 0.33 | 0.31 | 0.40 |
| Doc2 | 0.91 | 0.85 | 0.39 | 0.21  | 0.44 | 0.28 | 0.27 | 1.00 |
| Doc3 | 1.00 | 1.00 | 0.50 | 1.00  | 0.47 | 0.29 | 0.31 | 0.32 |
| Doc4 | 0.95 | 0.83 | 0.56 | 0.39  | 0.42 | 0.51 | 1.00 | 0.40 |
| Doc5 | 0.87 | 0.81 | 1.00 | 0.32  | 0.40 | 0.28 | 0.34 | 0.53 |
| Doc6 | 0.92 | 0.88 | 0.37 | 0.30  | 0.30 | 1.00 | 0.37 | 0.41 |

(d) $C_{Ppl,Loc}^{Freq}$

|      | Ppl1 | Ppl2 | Ppl3 | Ppl4 | Ppl5 | Ppl6 | Ppl7 | Ppl8 |
|------|------|------|------|------|------|------|------|------|
| Loc1 | 1.00 | 1.00 | 1.00 | 1.00 | 1.00 | 1.00 | 1.00 | 1.00 |
| Loc2 | 0.00 | 0.00 | 0.01 | 0.00 | 0.00 | 0.00 | 0.00 | 0.00 |
| Loc3 | 0.00 | 0.00 | 0.00 | 0.01 | 0.00 | 0.00 | 0.00 | 0.00 |
| Loc4 | 0.17 | 0.17 | 0.16 | 0.13 | 0.14 | 0.16 | 0.22 | 0.15 |
| Loc5 | 0.00 | 0.00 | 0.00 | 0.00 | 0.00 | 0.01 | 0.00 | 0.00 |
| Loc6 | 0.00 | 0.00 | 0.00 | 0.00 | 0.00 | 0.00 | 0.01 | 0.00 |
| Loc7 | 0.00 | 0.00 | 0.00 | 0.00 | 0.00 | 0.00 | 0.00 | 0.01 |

(e) $C_{Ppl,Ppl}^{Freq}$

|      | Ppl1 | Ppl2 | Ppl3 | Ppl4 | Ppl5 | Ppl6 | Ppl7 | Ppl8 |
|------|------|------|------|------|------|------|------|------|
| Ppl1 |      | 1.00 | 1.00 | 1.00 | 0.96 | 1.00 | 1.00 | 1.00 |
| Ppl2 | 1.00 |      | 0.94 | 0.90 | 1.00 | 0.88 | 0.93 | 0.95 |
| Ppl3 | 0.03 | 0.02 |      | 0.06 | 0.03 | 0.04 | 0.05 | 0.04 |
| Ppl4 | 0.02 | 0.02 | 0.04 |      | 0.03 | 0.03 | 0.04 | 0.02 |
| Ppl5 | 0.02 | 0.02 | 0.02 | 0.03 |      | 0.03 | 0.06 | 0.05 |
| Ppl6 | 0.02 | 0.02 | 0.03 | 0.03 | 0.03 |      | 0.06 | 0.03 |
| Ppl7 | 0.03 | 0.03 | 0.06 | 0.06 | 0.08 | 0.07 |      | 0.07 |
| Ppl8 | 0.02 | 0.02 | 0.03 | 0.02 | 0.05 | 0.03 | 0.05 |      |

Table 14. Normalized Duration-based Coupling Matrices

(a) $C_{Dev,Loc}^{Dur}$

|           | Device1 |
|-----------|---------|
| Location1 | 1.00    |
| Location2 | 0.00    |
| Location3 | 0.00    |
| Location4 | 0.17    |
| Location5 | 0.00    |
| Location6 | 0.00    |
| Location7 | 0.00    |

Table 14. Normalized Duration-based Coupling Matrices

(b) $C_{Doc,Loc}^{Dur}$

|      | Doc1 | Doc2 | Doc3 | Doc4 | Doc5 | Doc6 |
|------|------|------|------|------|------|------|
| Loc1 | 1.00 | 1.00 | 1.00 | 1.00 | 1.00 | 1.00 |
| Loc2 | 0.00 | 0.00 | 0.00 | 0.00 | 0.00 | 0.00 |
| Loc3 | 0.00 | 0.00 | 0.00 | 0.00 | 0.00 | 0.00 |
| Loc4 | 0.15 | 0.17 | 0.14 | 0.21 | 0.18 | 0.16 |
| Loc5 | 0.00 | 0.00 | 0.00 | 0.00 | 0.00 | 0.00 |
| Loc6 | 0.00 | 0.00 | 0.00 | 0.00 | 0.00 | 0.00 |
| Loc7 | 0.00 | 0.00 | 0.00 | 0.00 | 0.00 | 0.00 |

(c) $C_{Ppl,Doc}^{Dur}$

|      | Ppl1 | Ppl2 | Ppl3 | Ppl4 | Ppl5 | Ppl6 | Ppl7 | Ppl8 |
|------|------|------|------|------|------|------|------|------|
| Doc1 | 0.78 | 0.80 | 0.29 | 0.29 | 1.00 | 0.32 | 0.27 | 0.38 |
| Doc2 | 0.84 | 0.85 | 0.37 | 0.18 | 0.42 | 0.26 | 0.24 | 1.00 |
| Doc3 | 1.00 | 1.00 | 0.49 | 1.00 | 0.43 | 0.28 | 0.27 | 0.31 |
| Doc4 | 0.84 | 0.83 | 0.53 | 0.34 | 0.41 | 0.49 | 1.00 | 0.41 |
| Doc5 | 0.83 | 0.81 | 1.00 | 0.30 | 0.36 | 0.28 | 0.29 | 0.49 |
| Doc6 | 0.90 | 0.88 | 0.35 | 0.26 | 0.29 | 1.00 | 0.33 | 0.40 |

(d) $C_{Ppl,Loc}^{Dur}$

|      | Ppl1 | Ppl2 | Ppl3 | Ppl4 | Ppl5 | Ppl6 | Ppl7 | Ppl8 |
|------|------|------|------|------|------|------|------|------|
| Loc1 | 1.00 | 1.00 | 0.04 | 0.06 | 0.06 | 0.05 | 0.03 | 0.05 |
| Loc2 | 0.00 | 0.00 | 0.03 | 0.00 | 0.00 | 0.00 | 0.00 | 0.00 |
| Loc3 | 0.00 | 0.00 | 0.00 | 0.03 | 0.00 | 0.00 | 0.00 | 0.00 |
| Loc4 | 0.17 | 0.16 | 1.00 | 1.00 | 1.00 | 1.00 | 1.00 | 1.00 |
| Loc5 | 0.00 | 0.00 | 0.00 | 0.00 | 0.00 | 0.04 | 0.00 | 0.00 |
| Loc6 | 0.00 | 0.00 | 0.00 | 0.00 | 0.00 | 0.00 | 0.01 | 0.00 |
| Loc7 | 0.00 | 0.00 | 0.00 | 0.00 | 0.00 | 0.00 | 0.00 | 0.01 |

(e) $C_{Ppl,Ppl}^{Dur}$

|      | Ppl1 | Ppl2 | Ppl3 | Ppl4 | Ppl5 | Ppl6 | Ppl7 | Ppl8 |
|------|------|------|------|------|------|------|------|------|
| Ppl1 |      | 0.64 | 0.38 | 0.48 | 0.48 | 0.35 | 0.45 | 0.42 |
| Ppl2 | 0.59 |      | 0.34 | 0.42 | 0.47 | 0.31 | 0.41 | 0.38 |
| Ppl3 | 0.83 | 0.81 |      | 0.84 | 0.40 | 0.58 | 1.00 | 0.91 |
| Ppl4 | 0.55 | 0.53 | 0.44 |      | 0.72 | 0.15 | 0.52 | 0.16 |
| Ppl5 | 0.66 | 0.71 | 0.25 | 0.85 |      | 0.40 | 0.62 | 0.54 |
| Ppl6 | 0.70 | 0.70 | 0.53 | 0.26 | 0.58 |      | 0.92 | 0.66 |
| Ppl7 | 1.00 | 1.00 | 1.00 | 1.00 | 1.00 | 1.00 |      | 1.00 |
| Ppl8 | 0.66 | 0.65 | 0.65 | 0.22 | 0.62 | 0.52 | 0.72 |      |